\documentclass[12pt]{article}
\usepackage[polish,english]{babel}
\usepackage[latin1]{inputenc}
\usepackage{graphicx}
\usepackage{times}
\usepackage[T1]{fontenc}
\usepackage{epsfig}
\usepackage{amsmath}
\usepackage{empheq}
\usepackage{framed}
\usepackage{cancel}
\usepackage{diagbox}
\usepackage{booktabs}
\usepackage{cite}

\def\br{\begin{eqnarray}}
\def\er{\end{eqnarray}}
\def\be{\begin{equation}}
\def\ee{\end{equation}}
\def\({\left(}
\def\){\right)}

\def\rlx{\relax\leavevmode}
\def\IR{\rlx\hbox{\rm I\kern-.18em R}}

\def\u2{\mid u\mid^2}

\begin{document}
\begin{titlepage}

\begin{center}
{\bf \Large Static multi-soliton solutions in the affine $su(N+1)$ Toda models}
\end{center}

\vspace{.5cm}

\thispagestyle{empty}

\begin{center}
{J. Costa de Faria,~   P. Klimas~ }

\small

\par \vskip .2in \noindent
Universidade Federal de Santa Catarina,\\ Trindade, 88040-900, 
Florian\'opolis, SC, Brazil

%\par \vskip .2in \noindent
%$^{(\dagger)}$~Department of Mathematical Sciences,\\
 %University of Durham, Durham DH1 3LE, U.K.

\end{center}

\begin{abstract}
We study some static multi-soliton configurations in the $su(N+1)$ Toda models. Such configurations exist for $N>1$. We construct explicitly a multi-soliton solution for any $N$ and study conditions for having such solutions. The number of static solitons is limited by the rank of the $su(N+1)$ Lie algebra.  We give some examples of non-static multi-soliton solutions with static components.
 
\end{abstract}

\end{titlepage}

%%%%%%%%%%%%%%%%%%%%%%%%%%%%%%%%%%%%%%%%%%%%%%%%%%%
\section{Introduction}

The integrable field-theoretic models are very rare, however, they are interesting due to theirs very special mathematical properties. An exceptional character of integrable models has its origin in an underlying infinite number of conservation laws. Such conservation laws make a model very special what manifests in dynamics of its solutions. The integrability concept is well understood in low-dimensional models.  An explicit construction of conserved quantities is possible thanks to so-called zero-curvature formulation \cite{Lax, Zakharow}, where the integrability condition is given by the Lax-Zakharov-Shabat equation. There are also some results for higher-dimensional integrable systems \cite{AFG, AGW}.  The first non-topological model of an integrable two-dimensional system has been proposed by  D. J. Kortweg and  G. de Vries \cite{KdV} in order to describe some solitary non-linear waves  observed by J. S. Russell \cite{Russell}. The Russel's solitary waves  have been called {\it solitons}. 
 
A contemporary meaning of the  word soliton is much wider than a simple description of a solitary wave on  shallow water. It stands for solutions in integrable field theories. The characteristic properties of solitons manifest in the process of scattering. It is a well known fact that two solitons scatter without a notable change of their shapes. The only effect of scattering is a famous phase shift which is observed as a dislocation of an actual trajectory of  out coming soliton comparing to the trajectory of  incoming  soliton which would not have had been scattered. In a typical scattering process the final state contains a scattered objects and a radiation manifesting as  small propagating waves. However, such radiation is not present in the integrable models due to an infinite number of conserved quantities. This is another and very special property of integrable systems.

One of the most popular relativistic integrable models is the sine-Gordon model, introduced to describe the isometric embeddings of surfaces with constant negative Gaussian curvature in the Euclidean three-dimensional space \cite{sG}. The sine-Gordon multi-soliton solutions are usually obtained applying Hirota's method. It is a well known fact that for the sine-Gordon model there is not a solution which describes two solitons at rest. In other words each two sine-Gordon solitons always repel or attract. This fact became clear from the form of the Hirota's function describing the multi-soliton solution. In order to get such solution one can use a vertex operator representation of the Kac-Moody algebras \cite{FZ} for the case of $sl(2)$ algebra. The problem we would like to face here is existence of static multi-soliton solutions in models based on higher $sl(N+1)$ algebras. For this reason we shall study the family of Toda models \cite{HM, BB, H1, H2, Evans, OTU1, OTU2, KO, ACFGZ, LOT}. Such models form a family of two-dimensional integrable field theories. Moreover, the sine-Gordon model is in fact a special case corresponding with the $su(2)$ Toda model. The Toda models play also important role in integrable string theories where some version of such models appear when analyzing propagation of strings in $D=4$ de Sitter sapacetime \cite{VS}. Recently it has been established relation between Bethe Ansatz equations and $A_n^{(1)}$ Toda field theories \cite{AD}. It suggest wide aplications of solitons of Toda theories in other branches of physics.

For simplicity reasons we shall consider the case of the algebra $su(N+1)$. It turns out that in the case $N>1$ the static multi-soliton solutions do really exist. The existence of such solutions has been mentioned in \cite{ACFGZ} in the context of two soliton solution and it has been also considered in \cite{FKZ}. In this paper we shall report on static solutions for general multi-soliton solutions in the affine $su(N+1)$ Toda models. The general multi-soliton solutions are known \cite{ZC} but they have not been studied in details in this context.

The paper is organized as follows. In the second section we shortly discuss the model. The third section is devoted to a construction of the general $M-$soliton solution for arbitrary $N$.  We use the Hirota's method in order to find tau functions describing multi-soliton solutions. In fourth section we introduce the canonical energy-momentum tensor in terms of Hirota's tau functions. In fifth section we obtain conditions to get a static two-soliton solution. From the plots of energy density one can conclude that many static (and non-static) solutions have singular energy densities. We show how to choose a free parameters in Hirota's  functions in order to get non-singular energy density. In this section we also report on relation between the static $M-$soliton solution and the rank of the Lie algebra $su(N+1)$ of the corresponding Toda model. In sixth section we study some non-static solutions with static components. 

%%%%%%%%%%%%%%%%%%%%%%%%%%%%%%%%%%%%%%%%%%%%%

\section{The model}
We shall consider the $su(N+1)$ affine Toda  model defined by the action funcional in $(1+1)-$dim Minkowski space-time with a metric $g_{\mu\nu}={\rm diag}(1,-1)$
\br
S[\phi]=\int_{\Omega} d^2x\left[\frac{1}{4}\partial_{\mu}\vec\phi\cdot\partial^{\mu}\vec\phi-2\sum_{j=0}^N\left(1-e^{i\vec\alpha_j\cdot\vec\phi}\right)\right]
\er
where $
\vec\phi=\sum_{a=1}^N\frac{2\vec\alpha_a}{\alpha_a^2}\phi_a
$ is given as a combination of $N$ complex scalar fields $\phi_a(t,x)$, where $\vec\alpha_a$ stands  for simple roots of the $su(N+1)$ Lie algebra characterized by the $N\times N$ Cartan matrix
\br
K_{ab}\equiv\frac{2\,\vec\alpha_a\cdot\vec\alpha_b}{\alpha_b^2}={\scriptsize\left(\begin{array}{rrrrrrr}
2 & -1 & 0 & 0 & \hdots& 0& 0\\
-1& 2  & -1 & 0 &\hdots& 0& 0\\
0 & -1 & 2 & -1&\hdots& 0& 0 \\
\vdots & \vdots & \vdots & \vdots& \vdots& \vdots& \vdots \\
0 & \hdots & 0 & 0 &-1 & 2 &-1\\
0 & 0  & 0 &\hdots & 0 & -1 & 2  \\
\end{array}\right)}
\er
and $\vec\alpha_0:=-\vec\psi$, where $\vec\psi$ is the highest root given as the sum of simple roots $\vec\psi=\sum_{a=1}^N\vec\alpha_a$. The symbol $\alpha_b^2$ stands for the expression $\vec\alpha_b\cdot\vec\alpha_b$.
The principle of stationary action leads to the Euler-Lagrange equations which have the form
\br
\partial_{\mu}\partial^{\mu}\vec\phi=4i\sum_{j=0}^N\vec\alpha_je^{i\vec\alpha_j\cdot\vec\phi}.\label{eom1}
\er
One can obtain the field equations involving $\phi_a$ taking a scalar product of (\ref{eom1}) with fundamental weights $\vec \lambda_a$ which satisfy $\frac{2\vec\lambda_a\cdot\vec\alpha_b}{\alpha_b^2}=\delta_{ab}$. It leads to the set of $N$ equations having the form 
\br
\partial_{\mu}\partial^{\mu}\phi_a=2i{\alpha_a^2}\left[e^{i\vec\alpha_a\cdot\vec\phi}-e^{i\vec\alpha_0\cdot\vec\phi}\right].
\er 
 The energy density (see the formula (\ref{emtensor})) has the form 
\br
T^{0}_{\,\,\,0}=\frac{1}{4}\left[\partial_t\vec\phi\cdot\partial_t\vec\phi+\partial_x\vec\phi\cdot\partial_x\vec\phi\right]+V\label{energydensity}
\er
where the potential 
\br
V=2\sum_{j=0}^N\left(1-e^{i\vec\alpha_j\cdot\vec\phi}\right)\label{potencial}
\er
is so-called the Toda's potential. The energy density vanishes on constant solutions of (\ref{eom1}) given by
$$
\vec\phi_{vac}=2\pi\sum_{b=1}^Nn_b\vec\lambda_b
$$
where $n_b$ are some integers and $\vec\lambda_b$ stands for fundamental weights. It follows that the model has infinitely many degenerated vacua and therefore there are many soliton solutions connecting these vacua.  Such solitons are characterized by topological charge obtained from a zero component of the topological current density 
$\vec j_{\mu}^T=-\frac{1}{2\pi}\varepsilon_{\mu\nu}\partial^{\nu}\vec \phi$ i.e. 
\br
\vec Q_{T}=\int_{-\infty}^{\infty}dx\vec j_0^T=\frac{1}{2\pi}\left[\vec\phi(\infty)-\vec\phi(-\infty)\right]
\er
%%%%%%%%%%%%%%%%%%%%%%%%%%%%%%%%%%%%%%%%%%%%%

\section{The soliton solutions}
In this section we obtain multi-soliton solutions using Hirota's method and adequating notation to our purposes. An alternative derivation is presented in \cite{ZC}.  For didactic reason  we shall do it step by step.  One can obtain such solution in more elegant way using a vertex operator approach \cite{OTU2}. It is convenient to work with light cone coordinates $x_{\pm}=x\pm t$, then (\ref{eom1}) became $\partial_{+}\partial_{-}\vec\phi+i\sum_{j=0}^N\vec\alpha_je^{i\vec\alpha_j\cdot\vec\phi}=0$.  Let us express the $N$ fields $\phi_a$ in terms of $N+1$ Hirota's $\tau$ functions. The ansatz has the form 
\br
\phi_a=i\ln{\frac{\tau_a}{\tau_0}}\label{ansatz}
\er
and therefore $\vec\phi=i\sum_{k=0}^N\frac{2\vec\alpha_k}{\alpha_k^2}\ln\tau_k$. Let us observe that a trivial solution $\phi_a=0$ corresponds to $\tau_j=const$ which can be set $const=1$ because only the ratio of functions $\tau_j$ is important. The ansatz (\ref{ansatz}) allows to substitute $N$ equations of motion by the set of $N+1$  equations 
$$
\partial_{+}\partial_{-}\ln\tau_j=1-\prod_{k=0}^N\tau_k^{-K_{jk}},
$$
where $K_{jk}=\frac{2\vec\alpha_j\cdot\vec\alpha_k}{\alpha_k^2}$ with $j,k=0,1,\ldots,N$, represent elements of the extended Cartan matrix 
\br
K_{jk}={\scriptsize\left(\begin{array}{rrrrrrr}
2 & -1 & 0 & 0 & \hdots& 0& -1\\
-1& 2  & -1 & 0 &\hdots& 0& 0\\
0 & -1 & 2 & -1&\hdots& 0& 0 \\
\vdots & \vdots & \vdots & \vdots& \vdots& \vdots& \vdots \\
0 & \hdots & 0 & 0 &-1 & 2 &-1\\
-1 & 0  & 0 &\hdots & 0 & -1 & 2  \\
\end{array}\right)}\label{extendCartan}.
\er
The constant $\tau_j=1$ is obviously a solution of the last set of equations. The Hirota's function satisfy $\tau_{j+N+1}=\tau_j$ and the last equation can be cast in the form
\br
\tau_j\partial_{+}\partial_{-}\tau_j-(\partial_+\tau_j)(\partial_-\tau_j)=\tau_j^2-\tau_{j-1}\tau_{j+1}.\label{Hirotaeq}
\er
We shall refer to equation (\ref{Hirotaeq}) as to Hirota's equation. Following the idea of Hirota's method we shall focus on solution given in terms of a series expansion
\br
\tau_k=1+\sum_{n}\varepsilon_n\tau_k^{(n)}+\sum_{n,m}\varepsilon_n\varepsilon_m\tau_k^{(n,m)}+\ldots
\er
where the range of indices $n$ and $m$ as well as the form of functions $\tau_k^{(n,m,\ldots )}$  shall be specified below. The expansion parameters $\varepsilon_n$ are some auxiliary constants which helps to control the order of expansion. In fact, they do not have to be small and in the final stage of construction all $\varepsilon_n$ can be set as unity. The order of truncation of the series expansion define the number of solitons described by the multi-soliton solution. We shall start with the one soliton solution and then generalise a solution it to multi-soliton case.

\subsection{The one-soliton solution}\label{n1}
The one soliton solution can be obtained in terms of $\tau_k$ having the form  
\br
\tau_k=1+\varepsilon_n v^{(n)}_ke^{\Gamma}\label{1sol}
\er
where $n$ is a fixed integer number, $v_{k+N+1}^{(n)}=v_{k}^{(n)}$ and $\Gamma$ is given as a linear combination of the light-cone coordinates $x_{\pm}$
\br
\Gamma=\gamma_+^{(n)}x_++\gamma_-^{(n)}x_-+\xi\label{Gamma1sol}
\er
where $\xi$ is an arbitrary (in general complex) constant. Plugging the form of solution (\ref{1sol})  to (\ref{Hirotaeq})  one gets two algebraic equations corresponding to orders $\varepsilon_n$ and $\varepsilon_n^2$ which has to be satisfied simultaneously 
\br
&&\sum_{k=0}^NK_{jk}v^{(n)}_k=\lambda_n v^{(n)}_j\label{cond1}\\
&&v^{(n)}_kv^{(n)}_k-v^{(n)}_{k-1}v^{(n)}_{k+1}=0 \label{cond2}
\er
where $\lambda_n\equiv\gamma_+^{(n)}\gamma_-^{(n)}$ and $K_{jk}$ is the extended Cartan matrix (\ref{extendCartan}). The equation (\ref{cond1}) is just an eigenproblem for the extended Cartan matrix and it has the form of the Chebyshev recurrence relation. A trigonometric representation of the Chebyshev polynomials gives the expression for eigenvalues 
\br
\lambda_n=4\,\sin^2\left(\frac{n\,\pi}{N+1}\right), \qquad n=0,1,\ldots,N.\label{eigenvalues}
\er
It follows that the eigenvalue $\lambda_0=0$ is not degenerate and it has a corresponding eigenvector $v^{(0)}_k=1$. Also for $N+1$ being and even number the eigenvalue $\lambda_{(N+1)/2}=4$ is not degenerate and corresponding  eigenvector is of the form $v^{((N+1)/2)}_k=(-1)^k$. In all remaining cases the eigenvalues are degenerate what can be seen from the relation $\lambda_n=\lambda_{N+1-n}$. The eigenvectors associated with $\lambda_n$ are given by
$e^{ i \frac{2\pi}{N+1}nk}$ and $e^{-i \frac{2\pi}{N+1}nk}$. We can put all the cases together taking eigenvectors in the form
\br
v^{(n)}_k=e^{ i \frac{2\pi}{N+1}nk}, \qquad n=0,1,\ldots,N.\label{eigenvectors}
\er
One can verify that the second condition (\ref{cond2}) is automatically satisfied by (\ref{eigenvectors}). Taking $\gamma^{(n)}_{\pm}=\sqrt{\lambda_n}\,z^{\pm1}$ where $z$ are some free (in general complex) parameters we obtain
\br
\Gamma(n;z)=\sqrt{\lambda_n}\left[z\, x_++\frac{x_-}{z}\right]+\xi .\nonumber
\er
For $n=0$ all $\tau_k$  take the same constant value and the solution reduces to trivial one $\phi_a=0$. For this reason one-soliton solutions are those with non-zero $n$ i.e. $n=1,2,\ldots,N$. It follows that there are $N$-species of one-soliton solutions. The species-$n$ are given by
\br
\tau_k=1+e^{ i \frac{2\pi k}{N+1}n}e^{\Gamma(z)}\label{1soliton}.
\er
Choosing $z$ to be some real numbers $z=\eta e^{-\alpha}$, where $\tanh{\alpha}=v$  and $\eta=\pm1$, one gets
$$
z=\eta\sqrt{\frac{1-v}{1+v}}
$$
what leads to
\br
\Gamma(n;z)=2\,\eta\sqrt{\lambda_n}\,\frac{x-v\,t-x_0}{\sqrt{1-v^2}}\qquad {\rm where}\qquad \xi=-\frac{2\eta\,\sqrt{\lambda_n}}{\sqrt{1-v^2}}\,x_0.
\er
Let us note that the real part $\xi_R$ of the free constant $\xi=\xi_R+i\xi_I$ has been chosen as non-zero one whereas $\xi_I=0$. In general, one can consider $\xi_I\neq 0$. It is especially important for  solutions with odd $N$. For instance, the sine-Gordon Hirota's function can be obtained from the $N=1$ Toda model taking $\xi_I=\frac{\pi}{2}$. It turns out that with the right choice of the imaginary part $\xi_I$ one can avoid some non-integrable singularities in the energy density. Such singularities have origin in local vanishing of some $\tau_{(N+1)/2}$ functions. In next part of the paper we shall discuss this question.

\subsection{The two-soliton solution}
The proper two soliton solutions consist on species labeled by $n_1, n_2\in 1,\ldots N$. There are $N(N+1)/2$ different solutions of this kind. The species $(n_1,n_2)$ are given in terms of Hirota functions
\br
\tau_k=1+\varepsilon_1v_k^{(n_1)}e^{\Gamma_{1}}+\varepsilon_2v_k^{(n_2)}e^{\Gamma_{2}}+\varepsilon_1\varepsilon_2\delta_ke^{\Gamma_{1}+\Gamma_{2}}.\label{2soliton}
\er
where $\Gamma_i$ depends on $z_i$ and  $n_i$ i.e.  $\Gamma_i\equiv\Gamma_i(n_i;z_i)$. One can wonder why terms proportional to $\varepsilon_1^2$ and $\varepsilon_2^2$ are absent in (\ref{2soliton}). The answer is that it can be seen  from analysis of expansion of Hirota equation in order $\varepsilon^2_i$ which leads to condition of vanishing of such terms.The form of a  vector $\delta_k$ follows from vanishing of coefficient in expansion of Hirota equation in order $\varepsilon_1\varepsilon_2$  and it is given by expression
$
\delta_k=a(n_1,n_2;z_{1},z_{2})\,e^{ i \frac{2\pi k}{N+1}(n_1+n_2)},
$
where {\it the interaction coefficients} $a(n_1,n_2;z_{1},z_{2})$ read
\br
a(n_1,n_2;z_{1},z_{2})=\frac{z^2_1-\Omega^{(-)}_{n_1n_2}\,z_1z_2+z_2^2}{z^2_1+\Omega^{(+)}_{n_1n_2}\,z_1z_2+z_2^2}.\label{int2}
\er
The symbol $\Omega^{(\pm)}_{n_1n_2}$ is given in terms of eigenvalues (\ref{eigenvalues}) and it reads
\br
\Omega^{(\pm)}_{n_1n_2}:=\frac{\lambda_{n_1}+\lambda_{n_2}-\lambda_{n_1\pm n_2}}{\sqrt{\lambda_{n_1}\lambda_{n_2}}}.\label{omega2}
\er
The  Hirota's equation is satisfied in all orders. Setting $\varepsilon_1=\varepsilon_2=1$ one gets the expression
\br
\tau_k=1+e^{ i \frac{2\pi k}{N+1}n_1}e^{\Gamma_{1}}+e^{ i \frac{2\pi k}{N+1}n_2}e^{\Gamma_{2}}+a(n_1,n_2;z_{1},z_{2})\,e^{ i \frac{2\pi k}{N+1}(n_1+n_2)}e^{\Gamma_{1}+\Gamma_{2}}\label{2soliton}
\er

\subsection{The three-soliton solution}
It is quite instructive to study a three soliton solution before discussing the most general case. For such a solution the indices $i$ and $n_i$ take values 
$$
i,j\in 1,2,3\qquad\qquad n_i\in1,2,\ldots,N.
$$
The Hirota's $\tau$-functions contain terms up to third order proportional to $\varepsilon_1\varepsilon_2\varepsilon_3$. Similarly as for two-soliton case there are present only terms containing products of different $\varepsilon_i$ i.e  terms proportional to $\varepsilon_1^2\varepsilon_2$, $\varepsilon_1^3$ etc. are absent. The expression which solves the Hirota's equations  (\ref{Hirotaeq}) in all orders of expansion is of the form
\br
\tau_k=1+\sum_{i=1}^3v_k^{(n_i)}e^{\Gamma_i}+\sum_{i< j}^3a_{ij}v_k^{(n_i)}v_k^{(n_j)}e^{\Gamma_i}e^{\Gamma_j}+\prod_{i<j}^3 a_{ij}\prod_{c=1}^3v_k^{(n_c)}e^{\Gamma_c}.\label{tau3}
\er
The interaction coefficients in the second order of expansion have the form 
\br
a_{ij}\equiv a(n_i,n_j;z_i,z_j)=\frac{z_i^2-\Omega^{(-)}_{n_in_j}z_iz_j+z_j^2}{z_i^2+\Omega^{(+)}_{n_in_j}z_iz_j+z_j^2}\label{int3}
\er
where 
\br
\Omega^{(\pm)}_{n_in_j}:=\frac{\lambda_{n_i}+\lambda_{n_j}-\lambda_{n_i\pm n_j}}{\sqrt{\lambda_{n_i}\lambda_{n_j}}}.\label{omega}
\er
The highest-order term does not contain new type of coefficients. It depends on the product of the interaction coefficients (\ref{int3}) i.e. $\prod_{i<j}^3 a_{ij}=a_{12}a_{13}a_{23}$ whose functional form has exactly the same form as for two-soliton solution. It follows that the three-soliton solution is given in terms of expressions which have already been obtained for the two-soliton solution. The only difference is that there are three free parameters $z_i$ in this case. 

\subsection{The $M$-soliton solution}
It turns out that solutions with higher number of solitons do not contain  new kind of coefficients and therefore they can be expressed in terms of (\ref{int3}) and (\ref{omega}).
For the proper $M$-soliton solution the indices take values
$$
i\in 1,2,\ldots,M\qquad\qquad n_{i}\in1,2,\ldots,N
$$
In order to put a solution in compact form we shall extend the definition of the symbol $\prod_{i<j}^M a_{ij}$ to the case $M=1$ assuming that the product is defined as unity $\prod_{i<j}^1a_{ij}\equiv 1$  for $M=1$ and and it takes the usual form  
$$
\prod_{i<j}^M a_{ij}\equiv (a_{12}a_{13}\ldots a_{1M})(a_{23}a_{24}\ldots a_{2M})\ldots(a_{M-2\,M-1}a_{M-2\,M})(a_{M-1\,M})
$$
for $M>1$ .
The Hirota's $\tau$ functions take the form
\br
\tau_k&=&1+\sum_{S=1}^M\sum_{i_1< i_2<\ldots<i_S}^M\prod_{a<b}^Sa_{i_ai_b}\prod_{c=1}^Sv_k^{(n_{i_c})}e^{\Gamma_{i_c}}\label{intM},
\er
where term with $S=1$ does not contain coefficients $a_{ij}$ and the highest order term $S=M$ reduces to the single product
\br
\sum_{i_1< i_2<\ldots<i_M}^M\prod_{a<b}^Ma_{i_ai_b}\prod_{c=1}^Mv_k^{(n_{i_c})}e^{\Gamma_{i_c}}\equiv \prod_{i<j}^M a_{ij}\prod_{c=1}^Mv_k^{(n_c)}e^{\Gamma_c}.\label{lastterm}
\er 
Each summand with $S=K$ in (\ref{intM}) contains product of $K-$th exponents $e^{\Gamma_c}$. The number of interaction coefficients $a_{ij}$ in the product ("power" of $a$) equals to a number of essentially different pairs in the $K$-th element set i.e. $\binom{K}{2}=\frac{K(K-1)}{2}$. The number of summands in each multiple sum $\sum_{i_1< i_2<\ldots<i_K}^M$ equals to number of essentially different $K-$th element subsets in the $M-$th element set i.e. $\binom{M}{K}=\frac{M!}{K!(M-K)!}$.

%%%%%%%%%%%%%%%%%%%%%%%%%%%%%%%%%%%%%%%%%%%%%%%%%%

\section{The energy and the momentum densities}\label{n2}

The canonical energy-momentum tensor $T^{\mu}_{\,\,\,\,\nu}$ can be obtained directly from the form of Noether's currents corresponding to the space-time translation $x^{\mu}\rightarrow x'^{\mu}=x^{\mu}+\varepsilon^{\mu}$ i.e.
$$
j^{\mu}_{\alpha}=T^{\mu}_{\,\,\,\,\nu}\xi^{\nu}_{\alpha}
$$
where $\xi^{\nu}_{\alpha}:=\left.\frac{\partial x'^{\nu}}{\partial \varepsilon^{\alpha}}\right|_{\varepsilon=0}=\delta^{\nu}_{\alpha}$ are the corresponding Killing vectors. The Noether's currents have the general form \cite{arodzhadasz}
$$
j^{\mu}_{\alpha}=K^{\mu}_{\alpha}(\phi_a; x)-\sum_{b=1}^N\frac{\partial\mathcal{L}}{\partial(\partial_{\mu}\phi_b)}\mathcal{D}_{\alpha}\phi_b-\mathcal{L}\,\xi^{\mu}_{\alpha},
$$
where the term $K^{\mu}_{\alpha}(\phi_a; x)=\left.\frac{\partial K^{\mu}(\phi_{a};\, x;\,\varepsilon)}{\partial{\varepsilon^{\alpha}}}\right|_{\varepsilon=0}$ has origin in a surface term which is permissible in symmetry tansformation $S_{\Omega'}[\phi']=S_{\Omega}[\phi]+\int_{\partial\Omega}dS_{\mu}K^{\mu}(\phi_{a};\, x;\,\varepsilon)$\footnote{In quantum theories such term can be related to a change of phase factor of state vectors.}. We shall skip this term in our derivation of the energy-momentum tensor. The symbol $\mathcal{D}_{\alpha}\phi_b$ stands for Lie derivative of scalar field given by
$\mathcal{D}_{\alpha}\phi_b=-\xi^{\nu}_{\alpha}\partial_{\nu}\phi_b$. The Noether's currents are conserved for field $\phi_a$ being solutions of the Euler-Lagrange equations (\ref{eom1}).  In absence of $K^{\mu}_{\alpha}$ they have the form
$$
j^{\mu}_{\alpha}=\left[\sum_{b=1}^N\frac{\partial\mathcal{L}}{\partial(\partial_{\mu}\phi_b)}\partial_{\nu}\phi_b-\delta^{\mu}_{\nu}\mathcal{L}\right]\xi^{\nu}_{\alpha}
$$
and it follows that the conserved energy-momentum tensor reads
\br
T^{\mu}_{\,\,\,\,\nu}=\frac{1}{2}\partial^{\mu}\vec\phi\cdot\partial_{\nu}\vec\phi-\delta^{\mu}_{\nu}\left(\frac{1}{4}\partial^{\lambda}\vec\phi\cdot\partial_{\lambda}\vec\phi-V\right).\label{emtensor}
\er
The  components of the energy-momentum tensor have the form
\br
T^{0}_{\,\,0}&=&+\frac{1}{4}\left[(\partial_0\vec\phi)^2+(\partial_1\vec\phi)^2\right]+V\nonumber\\
T^{1}_{\,\,0}&=&-\frac{1}{2}(\partial_0\vec\phi)\cdot(\partial_1\vec\phi)\nonumber\\
T^{1}_{\,\,1}&=&-\frac{1}{4}\left[(\partial_0\vec\phi)^2+(\partial_1\vec\phi)^2\right]+V\nonumber
\er
We shall express the components of $T^{\mu}_{\,\,\nu}$ in terms of Hirota's $\tau$ functions.
The fields $\phi_a$ are given by 
$$
\phi_a=i\ln\frac{\tau_a}{\tau_0}
$$
Following \cite{LOT} we introduce the fields $\varphi_j$, $j=0,1,\ldots,N$ such that
$$
\vec\phi=\sum_{a=1}^N\frac{2\vec\alpha_a}{\alpha_a^2}\phi_a=\sum_{j=0}^N\vec e_{j}\varphi_j
$$
where $\vec e_i$ are some Cartesian versors 
\br
\vec e_i\cdot\vec e_j=\delta_{ij}\qquad\qquad\vec\alpha_j=\vec e_j-\vec e_{j+1}.
\er
It follows from this that 
$$
\vec\phi\cdot\vec\phi=\sum_{i=0}^N\sum_{j=0}^N\varphi_i\varphi_j(\vec e_i\cdot\vec e_j)=\sum_{j=0}^N\varphi_j^2.
$$
In terms of Hirota's functions $\tau_j$ the fields $\varphi_j$ read
\br
\varphi_j=i\ln\frac{\tau_j}{\tau_{j-1}}
\er
and therefore 
$$\partial_{\mu}\varphi_j=i\left(\frac{\partial_{\mu}\tau_j}{\tau_j}-\frac{\partial_{\mu}\tau_{j-1}}{\tau_{j-1}}\right).$$
Plugging this result to expressions for components of the energy-momentum tensor we get
\br
&&T^{0}_{\,\,\,0}=-\frac{1}{4}\sum_{j=0}^N\left[\(\frac{\partial_{0}\tau_j}{\tau_j}-\frac{\partial_{0}\tau_{j-1}}{\tau_{j-1}}\)^2+\(\frac{\partial_{1}\tau_j}{\tau_j}-\frac{\partial_{1}\tau_{j-1}}{\tau_{j-1}}\)^2\right]+V\nonumber\\
&&T^{1}_{\,\,\,0}=+\frac{1}{2}\sum_{j=0}^N\(\frac{\partial_{0}\tau_j}{\tau_j}-\frac{\partial_{0}\tau_{j-1}}{\tau_{j-1}}\)\(\frac{\partial_{1}\tau_j}{\tau_j}-\frac{\partial_{1}\tau_{j-1}}{\tau_{j-1}}\)\nonumber\\
&&T^{1}_{\,\,\,1}=+\frac{1}{4}\sum_{j=0}^N\left[\(\frac{\partial_{0}\tau_j}{\tau_j}-\frac{\partial_{0}\tau_{j-1}}{\tau_{j-1}}\)^2+\(\frac{\partial_{1}\tau_j}{\tau_j}-\frac{\partial_{1}\tau_{j-1}}{\tau_{j-1}}\)^2\right]+V\nonumber,
\er
where the potential (\ref{potencial}) is given by 
\br
V=2\sum_{j=0}^N\left(1-\frac{\tau_{j-1}\tau_{j+1}}{\tau^2_{j}}\right),
\er

%%%%%%%%%%%%%%%%%%%%%%%%%%%%%%%%%%%%%%%%%%%%%%%%%
\section{The static solutions}

\subsection{The static two-soliton solutions}
In this section we shall study conditions for some static two-soliton solutions. Let us observe that there are two special cases when $\lambda_{n_1}=\lambda_{n_2}\equiv\lambda_n$, namely
\begin{enumerate}
\item $n_1=n,$ $n_2=n$
\item $n_1=n$, $n_2=N+1-n$.
\end{enumerate}
It leads to significant simplification of the interaction coefficients (\ref{int2}). Then the construction of the static two-soliton solutions became quite natural.

\subsubsection{Case $n_1=n,$ $n_2=n$}
In this case the expressions (\ref{omega}) simplify to the form
$$
\Omega^{(-)}_{n_1n_2}=2-\frac{\lambda_0}{\lambda_n}=2\qquad\Omega^{(+)}_{n_1n_2}=2-\frac{\lambda_{2n}}{\lambda_n}
$$
where $\lambda_0\equiv0$. It follows that the interaction coefficients (\ref{int2}) take the form
\br
a(n,n;z_1,z_2)=\frac{(z_1-z_2)^2}{(z_1+z_2)^2-\frac{\lambda_{2n}}{\lambda_n}z_1z_2}.\label{case1}
\er
The interaction coefficient (\ref{case1}) vanishes for $z_1=z_2$ and as a consequence the two-soliton solution reduces to a one-soliton solution. Obviously, it does not lead to a static two soliton solution. However, when $z_1=-z_2$ the interaction coefficients became $a(n,n, z,-z)=4\frac{\lambda_{n}}{\lambda_{2n}}$. For odd $N$ the integer $n=\frac{N+1}{2}$ must be excluded because $\lambda_{N+1}=0$ what leads to singularity in $a$. 

The case $N=1$ is very special  because there is only one possible value $n=1$. The interaction coefficient has the form $a=(z_1-z_2)^2/(z_1+z_2)^2$. It is pretty clear from this expression that there are not static two-soliton solutions because the interaction coefficient is either zero or singular when $z_1=\pm z_2$. This fact is well known from the sine-Gordon model which in fact corresponds to $su(2)$ Toda model.

\begin{figure}[ht]
\begin{center}
\includegraphics[width=0.45\textwidth]{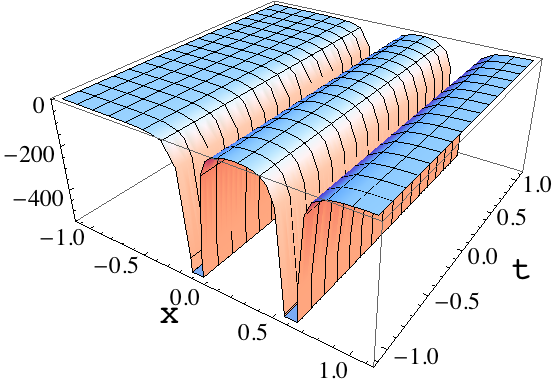}\hskip0.5cm\includegraphics[width=0.45\textwidth]{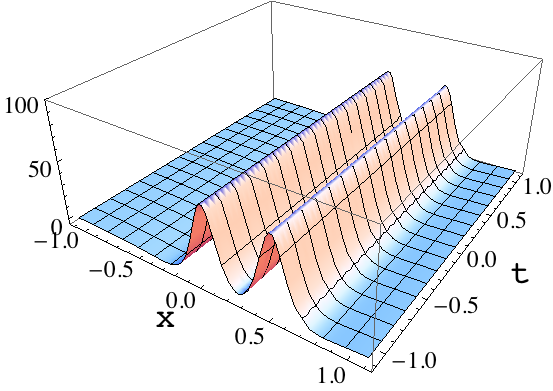}
\caption{The energy density for the $N=3$ static two-soliton solution for $(n_1,n_2)=(1,1)$, $(\eta_1,\eta_2)=(1,-1)$, $(v_1,v_2)=(0,0)$, $(x_0^1,x_0^2)=(1/2,0)$. Left picture: the energy density for $(\xi^1_I,\xi_I^2)=(0,0)$ has singular points, total energy diverges; right picture: the energy density for $(\xi^1_I,\xi_I^2)=(\pi/4,\pi/4)$ leads to a finite value of  the total energy.}\label{N3sing-nonsing}
\end{center}
\end{figure}

Except such special situations the solution has a form of a static two-soliton solution. Each two-soliton solution can be parametrized by velocities $v_j$ such that $z_j=\eta_j\sqrt{(1-v_j)/(1+v_j)}$. Such a solution is equivalent to a static one when $v_1=v_2$ because it can be transformed to an inertial frame where the solitons stay at rest. For this reason in our analysis of static solutions  we shall assume that $v_1=0=v_2$ what leads to
$\Gamma_j=2\eta_j\sqrt{\lambda_n}\,x+\xi_R^j+i\xi_I^j$. When both imaginary constants $\xi_I^1$ and $\xi_I^2$ vanish, one can conclude from (\ref{2soliton}) that Hirota's function satisfy $\tau_{N+1-k}=\tau_k^*$ and therefore they form some complex conjugated pairs. It leads to reality of the energy density. Unfortunately, for odd $N$ and vanishing  $\xi_I^j=0$  the energy density became singular for $n$ taking such values that $\frac{2nk}{N+1}$ is odd. It follows that a corresponding real-valued $\tau_k$ function vanishes at some $x$. In order to get a non-singular energy density one can choose the parameters $\xi_I^j$ as different from zero.  One can also preserve the reality of the energy density taking  $\xi_I^1=\xi_I^2\equiv \xi_I$ what leads to  
\br
\tau_k&=&1+\left[e^{\pm2\sqrt{\lambda_n}x+\xi_R^1}+e^{\mp2\sqrt{\lambda_n}x+\xi_R^2}\right]e^{i\left(\frac{2\pi k}{N+1}n+\xi_I\right)}\nonumber\\
&+&4\frac{\lambda_n}{\lambda_{2n}}e^{\xi_R^1+\xi_R^2}e^{2i\left(\frac{2\pi k}{N+1}n+\xi_I\right)}
\er 
The Hirota's functions form a complex conjugated pairs $\tau_{N+l-k}=\tau_k^*$, when the imaginary part takes the form $\xi_I=\pi\left[m-\frac{N+l}{N+1}n\right]$, where $l=0,1,\ldots,N$ represents all possible ways of conjugation and $m$ is some integer number. The integers $l$ and $m$ are not totally free because for
$\frac{2\pi k}{N+1}n+\xi_I$ being an odd integer number the related function $\tau_k$ vanishes for some $x$ what leads to singularities in energy density. All static two-soliton solutions for $n_1=n_2\equiv n$ are formed by species of the same kind. The Fig.\ref{N3sing-nonsing} shows the energy density for the $su(4)$ Toda model for two different choices of  imaginary constants $\xi_I^j$.

\begin{figure}[ht]
\begin{center}
\includegraphics[width=0.45\textwidth]{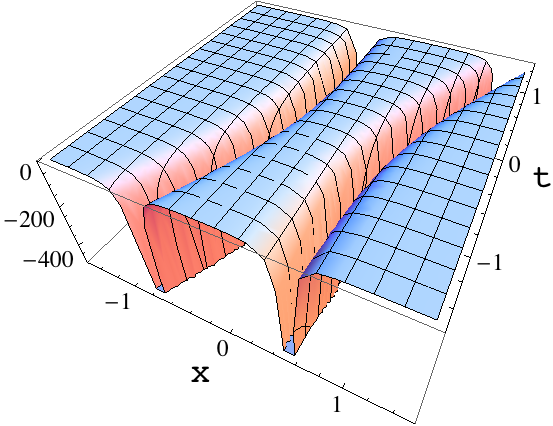}\hskip0.5cm\includegraphics[width=0.45\textwidth]{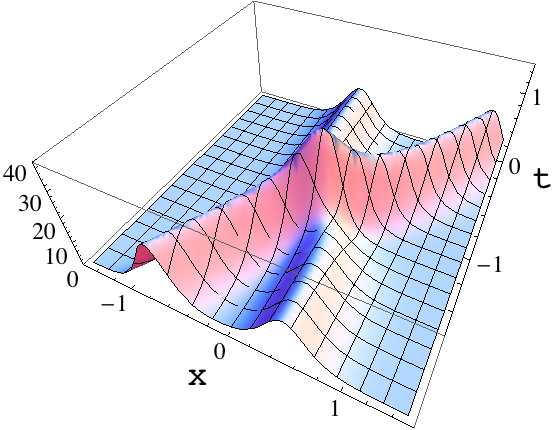}
\caption{The energy density for the $N=1$ two-soliton solution for $(\eta_1,\eta_2)=(1,-1)$, $(v_1,v_2)=(0,1/2)$, $(x_0^1,x_0^2)=(1/2,0)$. Left picture: $(\xi^1_I,\xi_I^2)=(0,0)$, right picture: $(\xi^1_I,\xi_I^2)=(\pi/2,\pi/2)$.}\label{N1sing-nonsing}
\end{center}
\end{figure}

Let us observe that also for some non-static soliton solutions one has to set $\xi_I\neq 0$ in order to get  a non-singular energy density. For instance, the energy density for $su(2)$ Toda model has singularities for $\xi_I=0$ and it behaves well for $\xi_I=\frac{\pi}{2}$, see Fig.\ref{N1sing-nonsing}. In fact for such a choice of parameters $\xi_I$ the Hirota's $\tau_k$ functions for $su(2)$ Toda model became exactly the Hirota's functions for the sine-Gordon model.

\subsubsection{Case $n_1=n,$ $n_2=N+1-n$}

For the second case  the coefficients (\ref{omega}) take the form
$$
\Omega^{(-)}_{n_1n_2}=2-\frac{\lambda_{2n}}{\lambda_n}\qquad\Omega^{(+)}_{n_1n_2}=2-\frac{\lambda_{N+1}}{\lambda_n}=2
$$
where $\lambda_{n-(N+1-n)}=\lambda_{2n-(N+1)}=\lambda_{2n}$ and  $\lambda_{n+(N+1-n)}=\lambda_{N+1}\equiv0$.
The corresponding interaction coefficients read
\br
a(n,N+1-n;z_1,z_2)=\frac{(z_1-z_2)^2+\frac{\lambda_{2n}}{\lambda_{n}}z_1z_2}{(z_1+z_2)^2}\label{case2}
\er
Similarly to the first case for $N=1$ there are not static two-soliton solutions because the interaction coefficient $a=(z_1-z_2)^2/(z_1+z_2)^2$ either vanishes or is singular for $z_1=\pm z_2$. The static solutions exist for $N=2,3,\ldots$ when $z_1=z_2$, however, one has to take care about the right choice of free constants in order to avoid singularities in energy density. In opposite to the previous case where eigenvectors $v_k^{(n_j)}$ were equal here they are mutually conjugated i.e. $v_k^{(n_1)}=e^{i\frac{2\pi nk}{N+1}}=(v^{(n_2)}_k)^*$. It suggest the choice of imaginary free constant as $\xi_I^1=-\xi_I^2\equiv \xi_I$. Then the Hirota's functions read
\br
\tau_k&=&1+e^{\pm2\sqrt{\lambda_n}x}\left[e^{\xi_R^1}e^{i\left(\frac{2\pi k}{N+1}n+\xi_I\right)}+e^{\xi_R^2}e^{-i\left(\frac{2\pi k}{N+1}n+\xi_I\right)}\right]\nonumber\\
&+&\frac{1}{4}\frac{\lambda_{2n}}{\lambda_{n}}e^{\pm4\sqrt{\lambda_n}x}e^{\xi_R^1+\xi_R^2}
\er 
Such functions form  complex conjugated pairs $\tau_{N+l-k}=\tau_k^*$, when the imaginary part takes the form $\xi_I=\pi\left[m-\frac{N+l}{N+1}n\right]$ where $l=0,1,\ldots,N$ and $m$ is an integer constant. Similarly to the previous case one has to avoid situation when $\frac{2kn}{N+1}+\frac{\xi_I}{\pi}$ is an odd integer number because it leads to $\exp\left[{i\left(\frac{2\pi k}{N+1}n+\xi_I\right)}\right]=-1$ and consequently corresponding $\tau_k$ vanish at some $x$. The static two-soliton with $n_1\neq n_2$ are formed by two solitons of different species. When $N$ is an odd number then for the special case $n_1=(N+1)/2$ also $n_2=(N+1)/2$ what leads to $\lambda_{2n}=0$. It follows that  $a=0$ in such a case and consequently instead of two-soliton solution one gets a one-soliton solution.

\begin{figure}[ht]
\begin{center}
\includegraphics[width=0.45\textwidth]{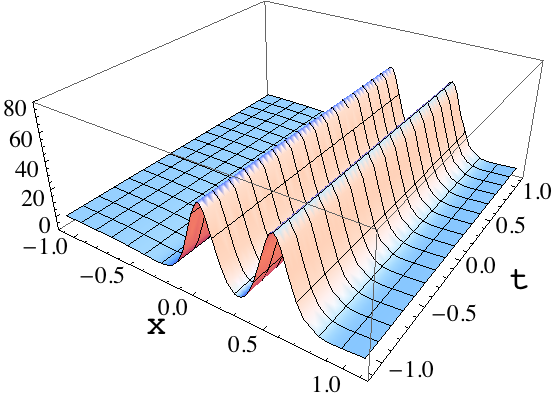}\hskip0.5cm\includegraphics[width=0.45\textwidth]{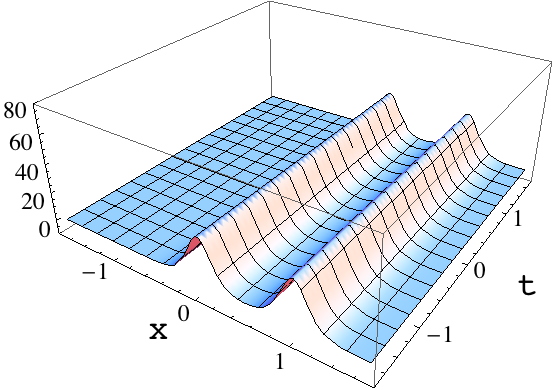}
\caption{The energy density for the $N=2$ static two-soliton solution for $(v_1,v_2)=(0,0)$, $(x_0^1,x_0^2)=(0.6,0)$, $(\xi^1_I,\xi_I^2)=(0,0)$. Left picture: the first case  $(n_1,n_2)=(1,1)$, $(\eta_1,\eta_2)=(1,-1)$; right picture: the second case $(n_1,n_2)=(1,2)$, $(\eta_1,\eta_2)=(1,1)$.}\label{N3twocases}
\end{center}
\end{figure}

The Fig.\ref{N3twocases} shows two static two-soliton solutions belonging to the first and to the second class. In this case both imaginary constants can be set $\xi_I^j=0$.

%%%%%%%%%%%%%%%%%%%%%%%%%%%%%%%%%%%%%%%%%%%%%%%%

\subsection{The static $M-$soliton solutions}

In this section we shall present some general statements about the static solutions in the affine $su(N+1)$ Toda model. The problem of singularities in the energy density is a very important question but it in fact is irrelevant for the problem of existence of static solutions. For this reason we shall not discuss this question here.

Let us consider a general $M-$soliton solution. This solution survives in static limit if all the interaction coefficients $a_{ij}$ which appear in the highest order term, proportional to $\prod_{i<j}^Ma_{ij}$, are neither zero nor infinity. There are  $\binom{M}{2}$ essentially different such coefficients. The indices $i,j$ run over the set $i,j\in\{1,2,\ldots,M\}$. For the case of the $su(N+1)$ Lie algebra the integers $n_i$ take values $n_i\in\{1,2,\ldots,N\}$ and it is  clear that for $N<M$ in the set $\{n_1,n_2,\ldots,n_M\}$ there are inevitable some repetitions.

Let us divide the values of the index $j$ in two sets $\cal{I}_-$ and $\cal{I}_+$, where for $j\in \cal{I}_-$ the sign parameter is negative $\eta_j=-1$  and similarly for $j\in \cal{I}_+$ it is positive $\eta_j=+1$. One can assume that the sets $\cal{I}_-$ and $\cal{I}_+$ have respectively $K$ and $M-K$ elements. In static limit the interaction coefficients have the form
\br
&&a_{ij}=\frac{2-\Omega^{(-)}_{n_in_j}}{2+\Omega^{(+)}_{n_in_j}}\qquad{\rm for}\qquad \eta_i\eta_j=1\label{a},\\
&&a_{ij}=\frac{2+\Omega^{(-)}_{n_in_j}}{2-\Omega^{(+)}_{n_in_j}}\qquad{\rm for}\qquad \eta_i\eta_j=-1\label{b}.
\er
We already know from previous considerations that $\Omega^{(-)}_{n_in_j}=2$ for $n_i=n_j$ and $\Omega^{(+)}_{n_in_j}=2$ for $n_i+n_j=N+1$. The problem of finding the static $M-$soliton solution reduces to the right choice of values of $n_i$ that guarantee that neither the coefficients (\ref{a}) became zero nor those given by (\ref{b}) became singular. The $M$ integers $n_i$ must be divided in two sets $\cal{N}_-$ and $\cal{N}_+$ in such a way that in none of them two integers $n_i$ repeat. Moreover, it must hold $n_i+n_j\neq N+1$, where $n_i\in \cal{N}_-$ and $n_j\in \cal{N}_+$. One can satisfy both these condition only if $N\ge M$. In such a case all values of $n_i$ can be chosen as different from each other. The first $K$ integers $n_i$ can be chosen freely from the $M-$th element set and associated with the set $\cal{N}_-$. Then one has to discard all integers $n_i$ in the $M-$th element set which satisfy $n_i+n_j=N+1$. Any of the remaining integers can be associated with the set $\cal{N}_+$. All possible combinations of integers which satisfy the conditions give the static $M-$soliton solutions.

The most important conclusion from this analyse is that the static $M-$soliton solution exist only for $N\ge M$.  This observation is in perfect concordance from the fact that the sine-Gordon model which is related to the $su(2)$ Toda model $(N=1)$ has not a static two soliton solution. For the same reason we never observe the static $M=3$ soliton solutions in the $su(3)$ Toda model. In other words, the existence of the static $M-$soliton solution require at least the $su(M+1)$ Lie algebra.

\begin{figure}[h!]
\begin{center}
\includegraphics[width=0.45\textwidth]{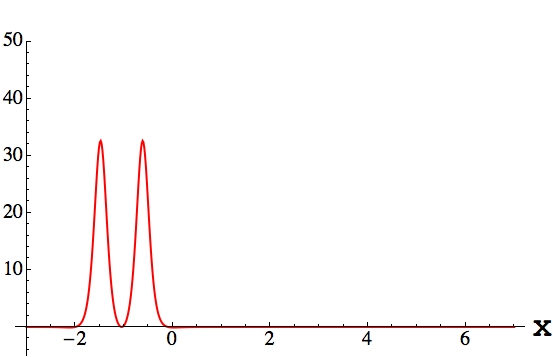}\includegraphics[width=0.45\textwidth]{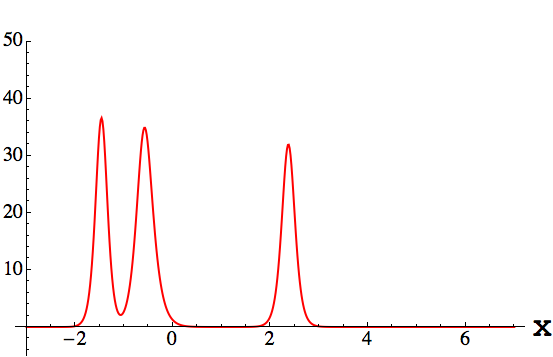}
\includegraphics[width=0.45\textwidth]{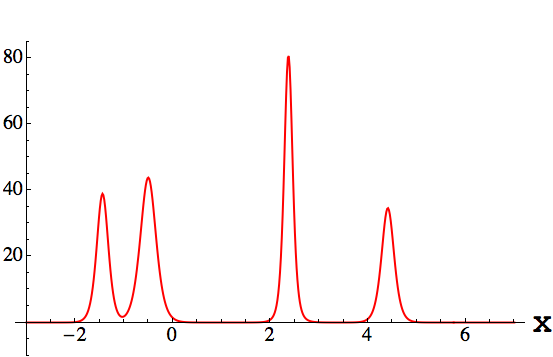}
\includegraphics[width=0.45\textwidth]{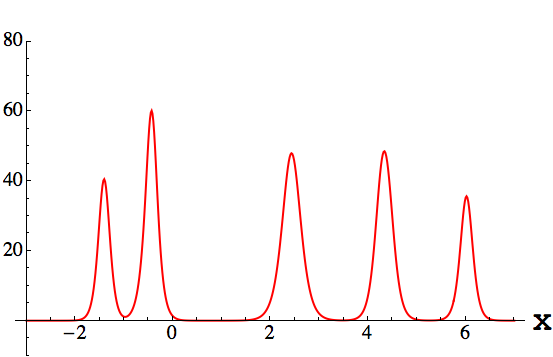}
\caption{The energy density for the static multi-soliton solutions of the $N=2,3,4,5$ Toda models. The number of solitons correspond with the value of $N$ and with the number of peaks.}\label{energy-plot}
\end{center}
\end{figure}

As the example we shall present some static multi-soliton solutions where the numer of solitons is exactly $N$. The highest five-soliton solution has been obtained for the $su(6)$ Toda model. We choose the case where all the sign parametars $\eta_i=1$. In such a case there is only a set ${\cal N}_+$.   We choose ${\cal N}_+=\{n_1=1,n_2=2,\ldots,n_N=N\}$. In order to get a real-valued energy density we set the imaginary free constants as $\xi_I^j=-\frac{N\pi}{N+1}n_j$. The real free constant have been chosen in terms of  $x_0^j$. Here $(x_0^1,x_0^2,x_0^3,x_0^4,x_0^5)=(-1.5,-1.0,1.5,3.0,4,0)$. The plots of the energy density are shown in Fig.\ref{energy-plot} whereas the total value of the total energy is presented in Tab.\ref{energy}. The energy grows faster than linearly with the number of solitons. 

\begin{table}[h]
\begin{center}
\begin{tabular}{|l||*{5}{c|}}\hline
\backslashbox{$ $}{$N$}
&\makebox[3em]{2}&\makebox[3em]{3}&\makebox[3em]{4}
&\makebox[3em]{5}\\\hline\hline
 $E$&$20.78$&$38.62$&$61.55$&$89.56$\\\hline
 $E/N$ &$10.39$&$12.87$&$15.38$&$17.91$\\\hline
\end{tabular}
\caption{The values of the total energy that corresponds with the energy density shown in Fig.\ref{energy-plot}.}\label{energy}
\end{center}
\end{table}

%%%%%%%%%%%%%%%%%%%%%%%%%%%%%%%%%%%%%%%

%%%%%%%%%%%%%%%%%%%%%%%%%%%%%%%%%%%%%%%

%%%%%%%%%%%%%%%%%%%%%%%%%%%%%%%%%%%%%%%%%%%%%%%
\section{Examples of the multi-soliton solutions with static components}

It turns out that some multi-soliton solutions which are not static could have some static component formed by two or more solitons which are at rest to each other. In this section we shall give some examples of such solutions. In order to avoid problems with singularities in energy density we shall study solutions of $su(N+1)$ Toda model for  $N=2,4,6$. We shall concentrate on some four-soliton configurations which are much more complex than two soliton-solution.

\begin{figure}[ht]
\begin{center}
\includegraphics[width=0.45\textwidth]{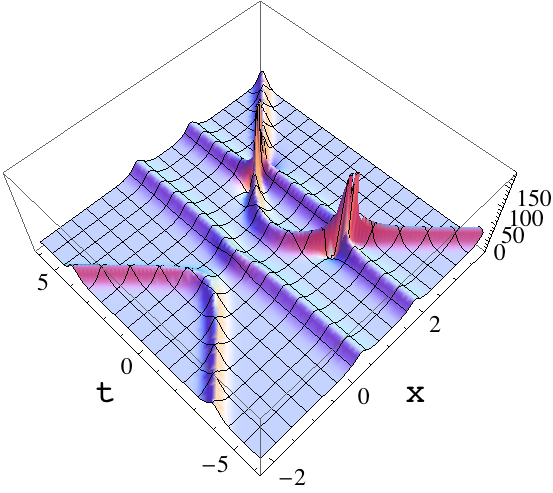}\hskip 0.5cm\includegraphics[width=0.45\textwidth]{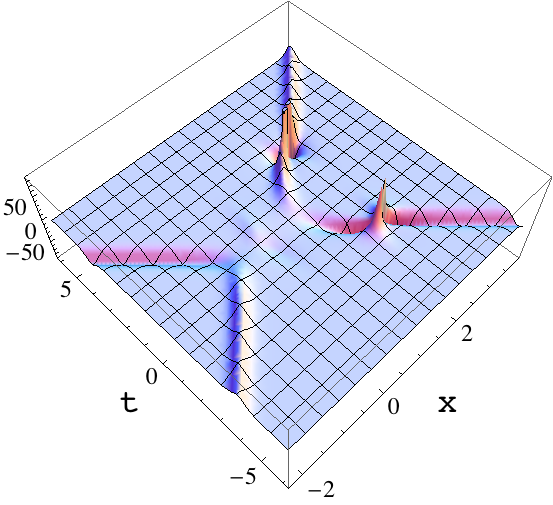}
\caption{The energy (left) and the momentum (right) densities for the $N=2$ four-soliton solution with a static component. The parameters of the solution: $(n_1,n_2,n_3,n_4)=(1,1,1,1)$, $(v_1,v_2,v_3,v_4)=(0.5,-0.5,0,0)$, $(x_0^1,x_0^2,x_0^3,x_0^4)=(0,0,1,0)$, $(\xi^1_I,\xi_I^2,\xi_I^3,\xi_I^4)=(0,0,0,0)$, $(\eta_1,\eta_2,\eta_3,\eta_4)=(1,1,-1,1)$.}\label{ex1-plot}
\end{center}
\end{figure}

As the first example we shall consider the four-soliton solution for the $su(3)$ Toda model. The solution is formed by four species of the first kind. We choose two velocity parameters as zero $v_3=v_4=0$ and two other as $v_1=1/2$ and $v_2=-1/2$. We shall not present explicit form for Hirota's functions because the expression is quite complex. The table Tab.\ref{ex1-tab} contains the list of interaction coefficients $a_{ij}$. For better transparency we present only half of the list of symmetric coefficients $a_{ij}=a_{ji}$. The coefficients with $i=j$ in fact do not enter to the formula (\ref{intM}) therefore all relevant interaction coefficients are non-zero. It means that the considered four-soliton solution do not reduce to some lower multi-soliton solution. The energy and the momentum density for the considered case is plotted in Fig.\ref{ex1-plot}. The picture clearly shows that the solution contains a static component which changes slightly only in the region of interaction.

\begin{table}[h!]
\begin{center}
\begin{tabular}{|l||*{5}{c|}}\hline
\backslashbox{$i$}{$j$}
&\makebox[3em]{1}&\makebox[3em]{2}&\makebox[3em]{3}
&\makebox[3em]{4}\\\hline\hline
1 &$0$&$\frac{4}{13}$&$\frac{2}{13} \left(11+6 \sqrt{3}\right)$&$\frac{2}{13} \left(11-6 \sqrt{3}\right)$\\\hline
2 &$a_{21}$&$0$&$\frac{2}{13} \left(11+6 \sqrt{3}\right)$&$\frac{2}{13} \left(11-6 \sqrt{3}\right)$\\\hline
3 &$a_{31}$&$a_{32}$&$0$&$4$\\\hline
4 &$a_{41}$&$a_{42}$&$a_{43}$&$0$\\\hline
\end{tabular}
\caption{The interaction coefficients $a_{ij}=a_{ji}$ for the solution plotted in Fig.\ref{ex1-plot}.}\label{ex1-tab}
\end{center}
\end{table}

%%%%%%%%%%%%%%%%%%%%%%%%%%%%%%%%%%%%%%

\begin{figure}[h!]
\begin{center}
\includegraphics[width=0.45\textwidth]{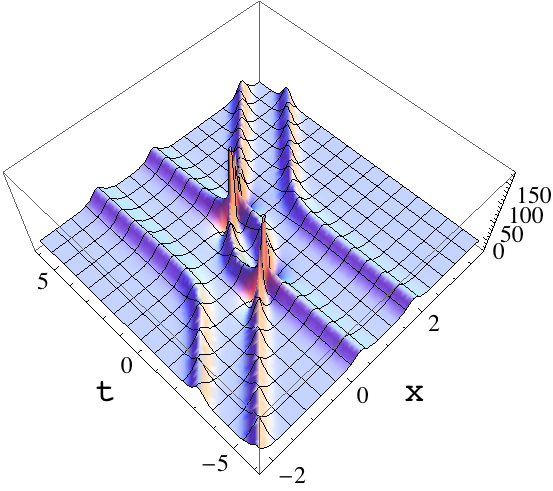}\hskip 0.5cm\includegraphics[width=0.45\textwidth]{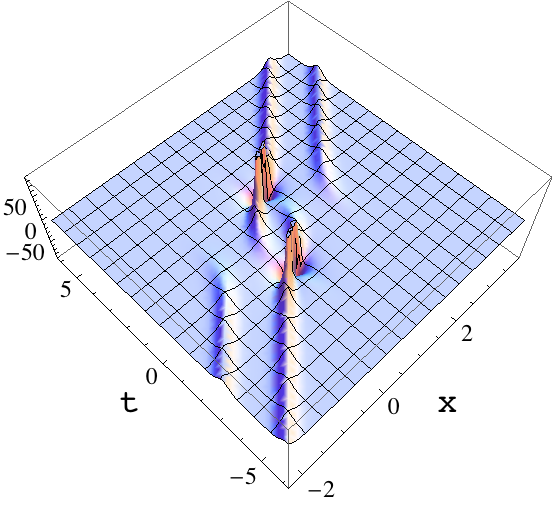}
\caption{The energy (left) and the momentum (right) densities for the $N=2$ four-soliton solution with two static components. The parameters of the solution: $(n_1,n_2,n_3,n_4)=(1,1,1,1)$, $(v_1,v_2,v_3,v_4)=(0.5,0.5,0,0)$, $(x_0^1,x_0^2,x_0^3,x_0^4)=(1,0,1,0)$, $(\xi^1_I,\xi_I^2,\xi_I^3,\xi_I^4)=(0,0,0,0)$, $(\eta_1,\eta_2,\eta_3,\eta_4)=(-1,1,-1,1)$.}\label{ex2-plot}
\end{center}
\end{figure}

As the second example we shall study the case with two static components. Taking $v_1=v_2$ and $v_3=v_3$ one can get a desired solution. In a chosen reference frame a velocity of one such component is zero whereas the second one takes value $v_1=v_2=1/2$. One can conclude from the table Tab.\ref{ex2-tab} that the solution is a full four-soliton (non-reduced) solution.  The figure Fig.\ref{ex2-plot} shows the behaviour of the maxima of the energy density in the scattering process. 
\begin{table}[h]
\begin{center}
\begin{tabular}{|l||*{5}{c|}}\hline
\backslashbox{$i$}{$j$}
&\makebox[3em]{1}&\makebox[3em]{2}&\makebox[3em]{3}
&\makebox[3em]{4}\\\hline\hline
1 &$0$&$4$&$\frac{2}{13} \left(11-6 \sqrt{3}\right)$&$\frac{2}{13} \left(11+6 \sqrt{3}\right)$\\\hline
2 &$a_{21}$&$0$&$\frac{2}{13} \left(11+6 \sqrt{3}\right)$&$\frac{2}{13} \left(11-6 \sqrt{3}\right)$\\\hline
3 &$a_{31}$&$a_{32}$&$0$&$4$\\\hline
4 &$a_{41}$&$a_{41}$&$a_{43}$&$0$\\\hline
\end{tabular}
\caption{The interaction coefficients $a_{ij}=a_{ji}$ for the solution shown in Fig.\ref{ex2-plot}.}\label{ex2-tab}
\end{center}
\end{table}

%%%%%%%%%%%%%%%%%%%%%%%%%%%%%%%%%%%%%%

\begin{figure}[h!]
\begin{center}
\includegraphics[width=0.45\textwidth]{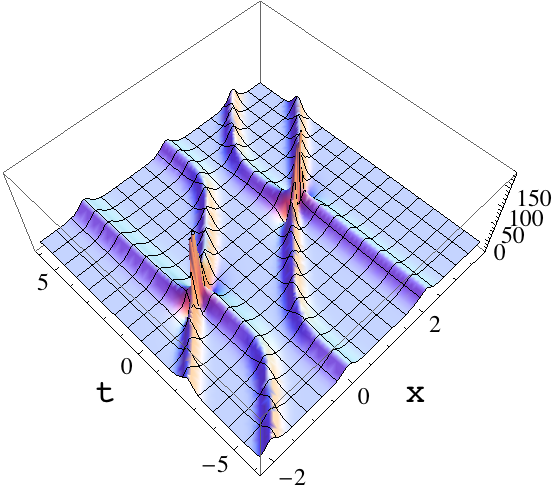}\hskip 0.5cm\includegraphics[width=0.45\textwidth]{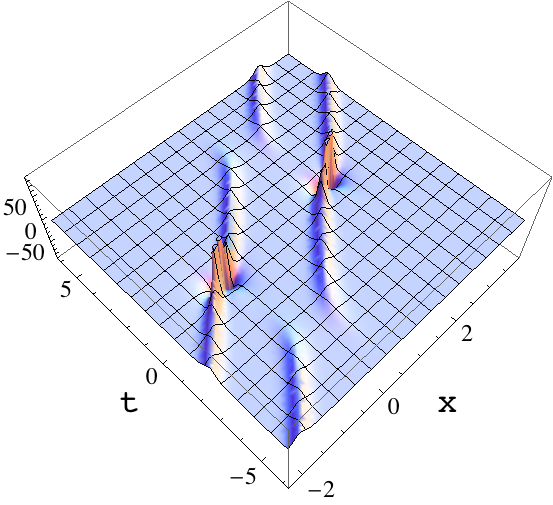}
\caption{The energy (left) and the momentum (right) densities for the $N=2$ four-soliton solution with two static components. The parameters of the solution: $(n_1,n_2,n_3,n_4)=(2,2,1,1)$, $(v_1,v_2,v_3,v_4)=(0.5,0.5,0,0)$, $(x_0^1,x_0^2,x_0^3,x_0^4)=(1,0,1,0)$, $(\xi^1_I,\xi_I^2,\xi_I^3,\xi_I^4)=(0,0,0,0)$, $(\eta_1,\eta_2,\eta_3,\eta_4)=(-1,1,-1,1)$.}\label{ex3-plot}
\end{center}
\end{figure}

In the third example we consider the four-solution solution for $su(3)$ Toda  which differs from the previous one only in species which form the solution. In the present case $n_1=n_2=2$ whereas $n_3=n_4=1$. The species of the second kind have velocity parameters $v_1=v_2=1/2$ and those of the first kind have $v_3=v_4=0$. Comparing Fig.\ref{ex2-plot} and  Fig.\ref{ex3-plot}  one can see a notable difference in behavior of the energy and momentum densities.

\begin{table}[h]
\begin{center}
\begin{tabular}{|l||*{5}{c|}}\hline
\backslashbox{$i$}{$j$}
&\makebox[3em]{1}&\makebox[3em]{2}&\makebox[3em]{3}
&\makebox[3em]{4}\\\hline\hline
1 &$0$&$4$&$\frac{11}{2}-3 \sqrt{3}$&$\frac{11}{2}+3 \sqrt{3}$\\\hline
2 &$a_{21}$&$0$&$\frac{11}{2}+3 \sqrt{3}$&$\frac{11}{2}-3 \sqrt{3}$\\\hline
3 &$a_{31}$&$a_{32}$&$0$&$4$\\\hline
4 &$a_{41}$&$a_{42}$&$a_{43}$&$0$\\\hline
\end{tabular}
\caption{The interaction coefficients $a_{ij}=a_{ji}$ for the solution shown in Fig.\ref{ex3-plot}.}\label{ex3-tab}
\end{center}
\end{table}

%%%%%%%%%%%%%%%%%%%%%%%%%%%%%%%%%%%%%%%

\begin{figure}[h!]
\begin{center}
\includegraphics[width=0.45\textwidth]{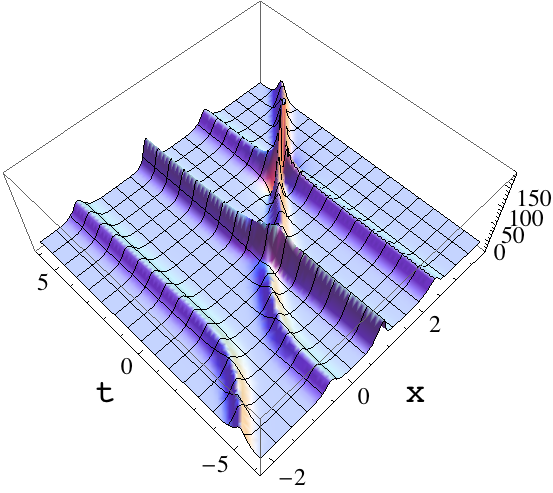}\hskip 0.5cm\includegraphics[width=0.45\textwidth]{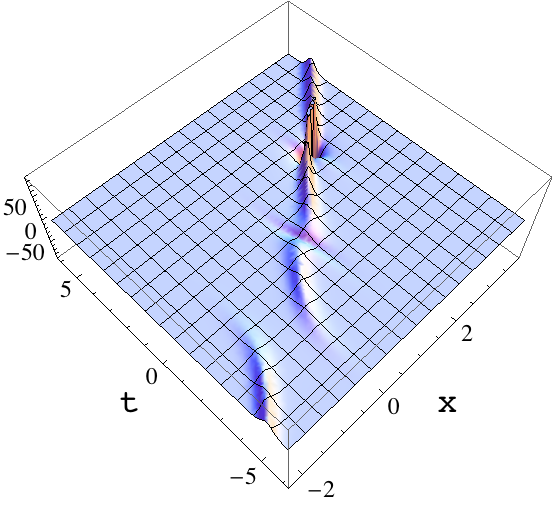}
\caption{The energy (left) and the momentum (right) densities for the $N=4$ four-soliton solution with a static component containing three maxima. The parameters of the solution: $(n_1,n_2,n_3,n_4)=(1,1,2,1)$, $(v_1,v_2,v_3,v_4)=(0,0.5,0,0)$, $(x_0^1,x_0^2,x_0^3,x_0^4)=(1.5,1.0,0.8,-0.5)$, $(\xi^1_I,\xi_I^2,\xi_I^3,\xi_I^4)=(0,0,0,0)$, $(\eta_1,\eta_2,\eta_3,\eta_4)=(1,-1,1,-1)$.}\label{ex4-plot}
\end{center}
\end{figure}

The multi-soliton solution for the Toda model with higher $N$ could have some static components with more than two maxima. As an example we shall study the solution for $su(5)$ Toda model which has only one velocity parameter being different from zero $v_2=1/2$. The list of interaction coefficients and the energy and the momentum densities are shown in  Tab.\ref{ex4-tab} and Fig.\ref{ex4-plot} respectively.

\begin{table}[h]
\begin{center}
\begin{tabular}{|l||*{5}{c|}}\hline
\backslashbox{$i$}{$j$}
&\makebox[3em]{1}&\makebox[3em]{2}&\makebox[3em]{3}
&\makebox[3em]{4}\\\hline\hline
1 &$0$&$\frac{2 \left(2+\sqrt{3}\right) \left(5-\sqrt{5}\right)}{20-5 \sqrt{3}-4 \sqrt{5}+3 \sqrt{15}}$&$\frac{4-\sqrt{6+2\sqrt{5}}}{4+\sqrt{6-2 \sqrt{5}}}$&$6-2 \sqrt{5}$\\\hline
2 &$a_{21}$&$0$&$\frac{8+\sqrt{18+6\sqrt{5}}}{8-\sqrt{18-6 \sqrt{5}}}$&$\frac{2 \left(2-\sqrt{3}\right) \left(5-\sqrt{5}\right)}{20+5 \sqrt{3}-4 \sqrt{5}-3 \sqrt{15}}$\\\hline
3 &$a_{31}$&$a_{32}$&$0$&$\frac{4+\sqrt{6+2\sqrt{5}}}{4-\sqrt{6-2 \sqrt{5}}}$\\\hline
4 &$a_{41}$&$a_{42}$&$a_{43}$&$0$\\\hline
\end{tabular}
\caption{The interaction coefficients $a_{ij}=a_{ji}$ for the four-soliton solution in the $su(5)$ Toda model shown in Fig.\ref{ex4-plot}.}\label{ex4-tab}
\end{center}
\end{table}

%%%%%%%%%%%%%%%%%%%%%%%%%%%%%%%%%%%%%%%

\begin{figure}[h!]
\begin{center}
\includegraphics[width=0.45\textwidth]{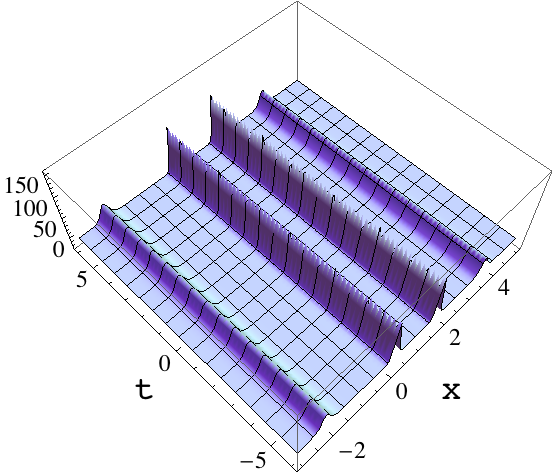}
\caption{The energy density for the $N=4$ static four-soliton solution. The parameters of the solution: $(n_1,n_2,n_3,n_4)=(1,2,2,1)$, $(v_1,v_2,v_3,v_4)=(0,0,0,0)$, $(x_0^1,x_0^2,x_0^3,x_0^4)=(-1.5,0.5,1.5,3.0)$, $(\xi^1_I,\xi_I^2,\xi_I^3,\xi_I^4)=(0,0,0,0)$, $(\eta_1,\eta_2,\eta_3,\eta_4)=(1,-1,1,-1)$.}\label{ex5-plot}
\end{center}
\end{figure}

\begin{figure}[h!]
\begin{center}
\includegraphics[width=0.45\textwidth]{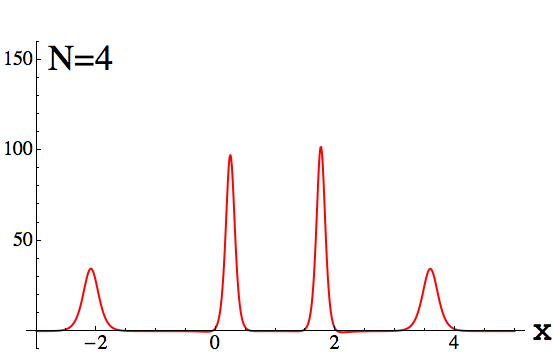}\includegraphics[width=0.45\textwidth]{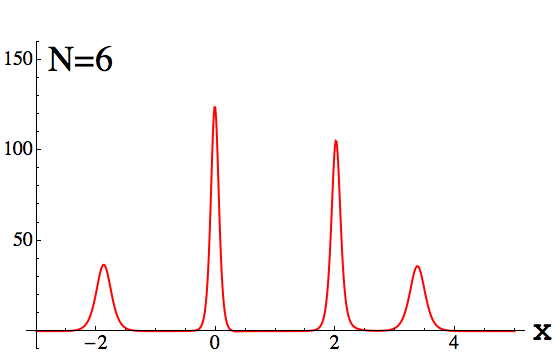}
\includegraphics[width=0.45\textwidth]{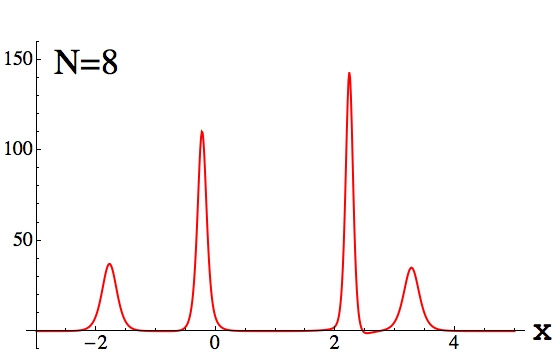}
\includegraphics[width=0.45\textwidth]{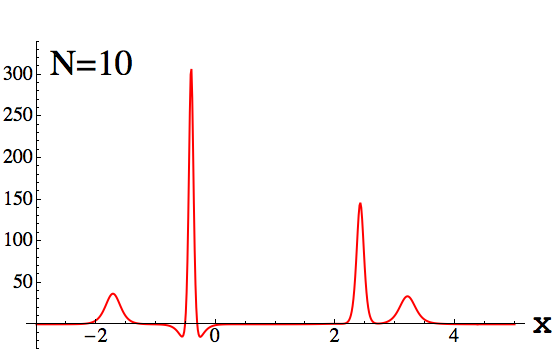}
\caption{The energy density for static four-soliton solution for the $su(5)$, $su(7)$, $su(9)$ and $su(11)$ Toda models. All solutions have the same values of parameters given in Fig.{\ref{ex5-plot}}}\label{ex6-plot}
\end{center}
\end{figure}

In all examples presented in this section some static configurations had been part of multi-soliton solutions. We already know that for $N=4$ there are static four-soliton solutions . The energy density of the static four-soliton solution is sketched in Fig.\ref{ex5-plot}. It has four maxima. The section of this picture for some fixed $t$ is shown on the first picture in Fig.\ref{ex6-plot}. The other pictures show energy density for Toda field with higher number of fields. The total energy grows with the number of fields and it take values: $E=61.55$ for $N=4$, $E=68.08$ for $N=6$, $E=70.91$ for $N=8$ and $E=72.37$ for $N=10$.

\begin{table}[h!]
\begin{center}
\begin{tabular}{|l||*{5}{c|}}\hline
\backslashbox{$i$}{$j$}
&\makebox[3em]{1}&\makebox[3em]{2}&\makebox[3em]{3}
&\makebox[3em]{4}\\\hline\hline
1 &$0$&$\frac{4+\sqrt{6+2\sqrt{5}}}{4-\sqrt{6-2 \sqrt{5}}}$&$\frac{4-\sqrt{6+2\sqrt{5}}}{4+\sqrt{6-2 \sqrt{5}}}$&$6-2 \sqrt{5}$\\\hline
2 &$a_{21}$&$0$&$6+2\sqrt{5}$&$\frac{4-\sqrt{6+2\sqrt{5}}}{4+\sqrt{6-2 \sqrt{5}}}$\\\hline
3 &$a_{31}$&$a_{32}$&$0$&$\frac{4+\sqrt{6+2\sqrt{5}}}{4-\sqrt{6-2 \sqrt{5}}}$\\\hline
4 &$a_{41}$&$a_{42}$&$a_{43}$&$0$\\\hline
\end{tabular}
\caption{The interaction coefficients $a_{ij}=a_{ji}$ for the four-soliton solution in the $su(5)$ Toda model shown in Fig.\ref{ex5-plot}.}\label{ex5-tab}
\end{center}
\end{table}

%%%%%%%%%%%%%%%%%%%%%%%%%%%%%%%%%%%%%%%

%%%%%%%%%%%%%%%%%%%%%%%%%%%%%%%%%%%%%%%

\section{Summary}

The study of static multi-soliton solution is based on the explicit expressions describing multi-soliton solutions in the affine $su(N+1)$ Toda model. We rederived such formula in a convinient notation.  The crucial point of the whole construction is the form of the interaction coefficients. The Hirota's tau functions contain products of such coefficients. This is quite obvious in the approach of the vertex operator. Analyzing the form of interaction coefficients  
in the static limit one can obtain necessary conditions for existence of static solutions. We have found that the integers $n_i$ cannot be totally arbitrary. Moreover, a given $M-$soliton solution does not disappear in static limit only if the rank of the Lie algebra is at least equal to $M$. For lower  $N$  there is not necessary number of all different integers $n_i$.

The second important observation is related with the energy density of the Toda solitons. We pointed out several cases when the energy density has some singularities. The origin of this fact is in the reality of some Hirota's tau functions. Such a situation is quite typical for odd $N$. The right choice of free complex parameters in the Hirota's functions allow to avoid situations when such functions vanish and the energy density is singular. It shows that the space of parameters of the Toda solitons is quite non-trivial because it has many holes. The solutions having parameter belonging to such holes are unacceptable.

We have studied explicitly only a particular parametrisation. It is pretty clear that a general non static multi-soliton solution allows to build up configurations containing also breathers \cite{HIG}. Having in mind static configurations we have not studied  such time-dependent solutions. Using our general expression for multi-soliton solution one can easily obtain such solutions. 

Finally it would be interesting to study a problem of static soliton solutions for some other Lie algebras. Especially one sholud answer a question about relation between a number of solitons and the rank of the Lie algebra of the Toda model.
 
\section{Acknowledgements}
The authors are indebted to Wojtek Zakrzewski, Nobuyuki Sawado and  Paulo Eduardo Goncalves  for disscusions. PK thanks to Andrzej Wereszczynski for helpful comments.

%%%%%%%%%%%%%%%%%%%%%%%%%%%%%%%%%%%%%%%

 \end{document}